# WSO/UV: World Space Observatory/Ultraviolet

M. Hernanz[a], R. González-Riestra[b], W. Wamsteker[c,d], B. Shustov[c,e], M. Barstow[c,f], N. Brosch[c,g], C. Fu-Zhen[c,h], M. Dennefeld[c,i], M. Dopita[c,j], A.I. Gómez de Castro[c,k], N. Kappelmann[c,l], I. Pagano[c,m], J. Sahade[c,n], H. Haubold[c,o], J.-E. Solheim[c,p], P. Martínez[c,q]

[a]IEEC/CSIC, Spain, [b]XMM SOC, VILSPA, Spain, [c]WIC: WSO Implementation Committee, [d]ESA-VILSPA, Spain, [e]INASAN, Russia, [f]University of Leicester, United Kingdom, [g]TAU, Israel, [h]USTC-CfA, China, [i]IAP, France, [j]ANU, Australia, [k]UCM, Spain, [l]IAAT, Germany, [m]CNR, Italy, [n]CONAE, Argentina, [o]UN-OOSA, Vienna, Austria, [p]University of Tromso, Norway, [q]SAAO, South Africa

**Abstract.** We summarize the capabilities of the World Space Observatory (UV) Project (WSO/UV). An example of the importance of this project (with a planned launch date of 2007/8) for the study of Novae is given.

## INTRODUCTION

The World Space Observatory/Ultraviolet (WSO/UV) project is a mission development initiated to overcome the severe lack of capabilities for UV observations for Astrophysics [1] in the planning time lines of the major Space Agencies (extending at least beyond the year 2014). Apart from its scientific importance, it represents a new mission implementation model for large space missions which can be applied to the need for large light collection power required to keep *space astrophysics* complementary to the continuously increasing sensitivity of *ground-based telescopes*. One of the assumptions, associated with the WSO concept is to avoid the excessive complexity required for multipurpose missions. Although there may exist purely technological or programmatic policy issues, which would suggest such complex missions to be more attractive, many other aspects, which do not need to be explored here, argue against such mission model. Following this precept, the first implementation model for a World Space Observatory has been done for the ultraviolet domain WSO/UV [2, 3, 4]. An innovative aspect of the WSO implementation model is the truly world-wide participation, already in the early stages of the project, generating fully independent access for astronomers from *all countries* through its open operations model.

## THE WSO/UV MISSION

The *primary mission* of the World Space Observatory is a spectroscopic capability extending over the wavelength range: $\Delta\lambda$ **103 - 310 nm.** Three individual spectrographs will be mounted at a distance of 50 mm from the optical axis of the telescope (T-170M) with an image quality of < 0.5 arcsec. Two resolution modes are supported: $R(\lambda/\delta\lambda) \approx 55,000$ for point source (Ø 1 arcsec), and $R(\lambda/\delta\lambda) \approx 1,000$ for Long Slit (1 x 70 arcsec). In the high resolution mode, the wavelength range has to be split in two sections to be able to match the spectrum on the detector (MCP) face. The long slit-low resolution mode is foreseen to cover the full range from $\Delta\lambda$ 103 - 310 nm in a single exposure.

As *secondary instrumentation* there will be a suite of 2 UV imagers with 6 filters (High Sensitivity Imager: **HSI** and High Resolution Imager: **HRI**) with redundancy; an optical imager (**OI**); and a special UV imaging device (**CFC**) placed at the optical axis of the telescope to utilize the high optical quality of the T-170M. In principle the operational mode will be an observing program driven by the spectrographic demand, with continued exposures for the imagers. Of course also specific imaging programs are foreseen, and will be supported. The field-of-view of the imagers, their wavelength coverage and focal ratio are indicated in table 1.

The *orbit* for WSO/UV is planned to be a halo type orbit around the second Earth-Sun Lagrangian point (L2).

The combined WSO system, which is based on current top-of-the-line technology (Technology Readiness Level TRL > 6); high throughput of the telescope/spectrograph; a limited number of primary instruments on board; and the orbit efficiency of the WSO/UV, will be such that an order of magnitude improvement will be obtained with respect to that associated with the Hubble Space Telescope. A very important reason for WSO/UV is supplied by the fact that, at its scheduled start of operations phase, the results of the GALEX UV survey will be available. No other mission is planned which would allow to use its results for further studies of the GALEX Catalogue, which will contain many objects which will be of great importance for the understanding of cosmological evolution of the overall content of the Universe.

## AN EXAMPLE OF WHAT WSO/UV WILL BE ABLE TO CONTRIBUTE TO NOVA STUDIES

To illustrate the importance of the observing capabilities supplied by the WSO/UV mission for the study of Novae, we show in Figure 1 the results obtained for Nova V1974 Cygni 1992 where, through the availability of the International Ultraviolet Explorer, a completely new insight in the evolution of the nova phenomenon was obtained, revolutionizing our understanding of the explosive processes driving a nova [5].

The capabilities of WSO will allow to obtain high resolution UV spectra of novae of similar luminosity in M31 with a signal-to-noise ratio of 10 in only 15 minutes. Even for novae in M81 (at 3.3 Mpc) useful spectra can be obtained with reasonable exposure times (of the order of 1-2 hours).

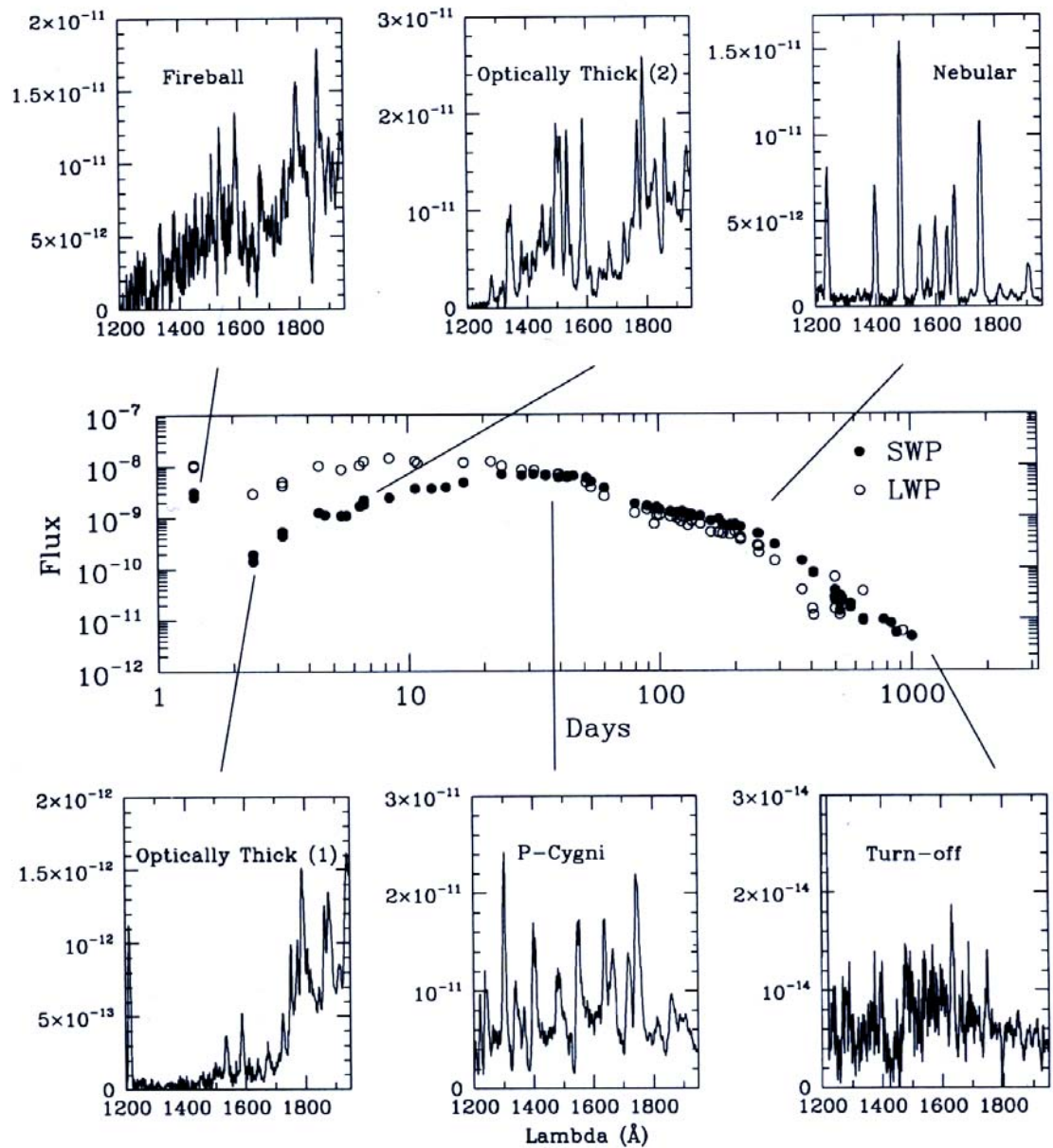

**FIGURE 1.** This figure shows the photometric and spectroscopic ultraviolet evolution of the bright Nova V1974 Cygni 1992 (from IUE data), the best observed nova so far over all the electromagnetic spectrum. The complete time coverage of the UV observations of this nova allowed to characterize the different phases of its evolution, even identifying some of them never observed before. Particularly well seen in this object is the "Constant Bolometric Luminosity Phase", predicted by the standard Thermonuclear Runaway Theory. The peak bolometric luminosity of this nova during this phase was of the order of $10^{38}$ erg s$^{-1}$.

**TABLE 1. Imaging with WSO/UV.**

| Camera | HRI (f/50) | HIS (f/10) | CFC (f/10) | OI (f/10/f/50) |
|---|---|---|---|---|
| $\Delta\lambda$ (nm) | $\lambda\lambda 115 - \lambda\lambda 310$ | | | $\lambda\lambda 360 - \lambda\lambda 800$ |
| F.O.V. (arcsec) | 60 | 300 | 300 | 240 |

[D80 ($\lambda$ 630 nm) = 0.35 arcsec]